\documentclass[11pt]{article}
\usepackage{latexsym}
\usepackage{amssymb}
\usepackage{epsf}
\usepackage{amsmath}
\usepackage{graphicx}

\newcommand{\lsim}{\lesssim}
\newcommand{\gsim}{\gtrsim}

\def\beq{\begin{eqnarray}}
\def\eeq{\end{eqnarray}}

\def\v{\varphi}
\def\D{\Delta}

\begin{document}
\pagestyle{empty}

\begin{flushright}
YITP-14-54 \\ 
\end{flushright}

\vspace{3cm}

\begin{center}

{\bf\LARGE  Fine Tunings for Inflation \\ \vskip.5cm with Simple Potentials}
\\

\vspace*{1.5cm}
{\large 
Izawa K.-I.
} \\
\vspace*{0.5cm}

{\it Yukawa Institute for Theoretical Physics, Kyoto University, Kyoto
 606-8502, Japan}\\
\vspace*{0.3cm}
{\it Kavli Institute for the Physics and Mathematics of the Universe (WPI),\\
 University of Tokyo, Kashiwa 277-8583, Japan}\\
\vspace*{0.5cm}

\end{center}

\vspace*{1.0cm}

\begin{abstract}
{
\pagestyle{plain}
We continue to investigate possible parameter choices
for primordial inflation with simple potentials
such as power-law and two-term potentials.
We examine the amount of parameter tuning
to make the slow-roll inflation eternal.
In particular, we adopt the critical coupling for marginally eternal inflation
and see whether realistic primordial inflation is attainable
under such a parameter choice.
It turns out that potentials with mass scale of grand unification
provide, after tuning, consistent results to experimental data so that
the considered tuning is allowed observationally.
Namely, the primordial inflation is possibly marginal in such a setup.
}
\end{abstract} 

\newpage
\baselineskip=18pt
\setcounter{page}{2}
\pagestyle{plain}
\baselineskip=18pt
\pagestyle{plain}

\setcounter{footnote}{0}

\section{Introduction}

Landscape of vacua can be regarded as a setup to answer
the question how the laws of nature are chosen among possible various theories.
In the landscape of vacua, background vacua with nearly flat potentials
around them are
expected to realize cosmic inflation which results in macroscopic
universes out of microscopic quantum fluctuations.
{}From a perspective of a vacuum so chosen, the potential
seems to be fine-tuned for flatness of potential. Let us make one-step
further assumption that the potential is tuned so that eternal inflation
is realized. Is such a physical hypothesis compatible
with the observational data?

It depends on representative models we choose whether eternal
inflation is consistent or inconsistent to the observational data.
For example, the model of the supersymmetric Higgs
inflation investigated from this viewpoint in
Ref.\cite{Imai}
implies that the spectral index is too small
for the case of eternal inflation, while minimal fine tuning
is concordant with experimental results.
This paper is partially motivated by this negative result on fine
tuning for eternal inflation and intended to search a possible realistic example
of such a tuning. 
The example we adopt is simpler than the model of supersymmetric
Higgs inflation, which has a particle physics concern of electroweak
hierarchy. In particular, we do not assume supersymmetry,
though it could be incorporated straightforwardly.

In this paper, we try to mimic GUT potentials with GUT scale mass to consider
inflaton potentials with GUT scale `mass' parameter.
This `mass' leads to effective intermediate scale after
fine tuning of over-all coupling constant.
We restrict ourselves to the simplest forms of inflaton potentials
such as power-law and two-term potentials
postponing investigations on more involved setups for various candidate inflation
scenarios
\cite{Martin:2013tda}.
Our present setup allows fine tunings for eternal inflation
that is compatible with recent experimental information such as the
Planck satellite and BICEP2 results
\cite{Ade:2013uln,BICEP2}.
This gives a concrete example of possible answer to the question how the model parameters are
chosen or tuned to realize appropriate slow-roll inflation.

The rest of the paper is organized as follows. In the next section,
we present one-parameter tuning for over-all coupling constant
with inflaton potentials of power-law form.
We recapitulate expressions for inflationary density fluctuations
and spectral indices in such a simple case for direct reference.
In section 3, we consider parameter tuning to realize eternal inflation
within the above setup. In particular, we adopt the critical coupling
for marginally eternal inflation
\cite{Izawa}
and argue that the resultant marginal
inflation is compatible with the observational constraints on
primordial inflation. 
In section 4, we comment on the case of two-term potential
through investigation of the polynomial potential similar to GUT potential.
The final section concludes the paper.

\section{The power-law term}

We take the simplest potential of monomial form as a starting point
to investigate parameter tuning for eternal inflation.

The Lagrangian is given by
\begin{eqnarray}
{\cal L}={1 \over 2\lambda^2}\partial_\mu \phi \partial^\mu \phi - {1 \over 2n}v^{4-2n} \phi^{2n},
\end{eqnarray}
where the real scalar field $\phi$ takes values in the range $|\phi|<a$
as the effective theory description is applicable in the present vacuum
in the landscape of vacua.
The constants $v$ and $a$ are presumably of order one or less
in the Planck unit
$M_G \simeq 2.4 \times 10^{18} {\rm GeV}=1$ and the overall coupling $\lambda$ is the parameter to be tuned
for inflation to take place within the regime $|\phi|<a$
around the specified vacuum.

Let us define a rescaled field $\varphi=\phi / \lambda$.
Then the kinetic term for the field $\varphi$ is canonical
and its potential is given by
\begin{eqnarray}
V(\varphi)={1 \over 2n}\lambda^{2n} v^{4-2n} \varphi^{2n},
\end{eqnarray}
with $|\varphi|<a / \lambda$.
This monomial form has no sound foundation for its selection
so that the present choice is just for simplicity of presentation.
{}From a more general perspective, we have assumed a situation where the single power-law
term is dominant when the parameter $\lambda$ is tuned.
The slow-roll inflationary dynamics of the present simple potential
is well known
\cite{Lyth}.
Note that our subject here is not to analyze the dynamics
itself but to examine parameter tuning to realize eternal inflation
in the present setup. For that purpose, let us recapitulate
the known analysis.

Hereafter we restrict ourselves to the regime $\varphi>0$
without loss of generality.
The slow-roll parameters are given by
\begin{eqnarray}
\epsilon(\varphi) &\equiv& \frac{1}{2}\Big(\frac{V'}{V}\Big)^2
 \simeq {1 \over 2}\left({2n \over \varphi}\right)^2, \label{eps} \\
\eta(\varphi) &\equiv& \frac{V''}{V}
 \simeq {2n(2n-1) \over \varphi^2}. \label{et}
\end{eqnarray} 
The absolute values of these parameters are smaller than one
during inflation.

The inflation ends when the slow-roll parameter $\epsilon$ or $|\eta|$ reaches one so that
the end point $\varphi=\varphi_f$ of slow-roll inflation is given by
\begin{eqnarray}
\varphi_f \simeq \sqrt{2}n, \label{end1}
\end{eqnarray} 
or
\begin{eqnarray}
\varphi_f \simeq |2n(2n-1)|^{1 \over 2}. \label{end2}
\end{eqnarray}
The e-fold number $\cal N$ is given by
\begin{eqnarray}
{\cal N}\simeq \int^{\varphi_{\cal N}}_{\varphi_f} \! d\varphi \ {V \over V'},
\end{eqnarray}
where $\varphi_{\cal N}$ denotes the field value of $\varphi$ corresponding to
the e-fold $\cal N$.
Hence it is obtained as
\begin{eqnarray}
\varphi_{\cal N}\simeq \sqrt{4n{\cal N} + \varphi_f^2} \simeq \sqrt{4n{\cal N}},
\end{eqnarray}
where we assume $\cal N$ of several tens and $0<n \leq 2$.

For direct reference, we here summarize resultant properties of the density fluctuations.
The parameters which characterize the primordial inflation model
are constrained by observational data.
In particular, we have restrictions on
the amplitude of density fluctuations
$\delta \rho/\rho$ corresponding to the horizon exit of the
present horizon with e-fold $\cal N$, which is given by
\begin{eqnarray}
{\delta \rho \over \rho}
 \simeq {1 \over 5\sqrt{3}\pi}
{V^{3 \over2}(\varphi_{\cal N}) \over |V'(\varphi_{\cal N})|}
 \simeq {1 \over 5\sqrt{3}\pi (2n)^{3 \over 2}}\lambda^n v^{2-n}
(4n{\cal N})^{{n+1 \over 2}}.
\label{density}
\end{eqnarray}
This should turn out to be the observed size of $2 \times 10^{-5}$
\cite{Ade:2013uln}.
Then the spectral index $n_s$ is obtained as
\begin{eqnarray}
n_s \simeq 1-6\epsilon(\varphi_{\cal N})+2\eta(\varphi_{\cal N}))
\simeq 1-{n+1 \over {\cal N}},
\end{eqnarray}
and the tensor-to-scalar ratio $r$ is given by
\begin{eqnarray}
r \simeq 16\epsilon(\varphi_{\cal N}) \simeq 8{n \over {\cal N}},
\end{eqnarray}
as is well known.
The observational data at present seems to suggest $n \lsim 1$
\cite{Ade:2013uln,BICEP2}.

\section{Inflationary parameter tuning}

We now consider possible criteria
for inflationary parameter tuning and examine whether such criteria are
realistic or not.
In particular,
we pursue parameter tuning
to make slow-roll inflation `marginally' eternal.
Namely, we consider marginal inflation
\cite{Izawa},
where less tuned potential only realizes non-eternal inflation,
or `irrelevant' inflation,
and more tuning leads to `relevant' inflation, not to say eternal.
The nomenclature is borrowed from that in
renormalization theory such as (non-)renormalizable interactions,
or (ir)relevant ones.

Realization of slow-roll inflation
implies parameter tuning of an inflaton potential
in order to make it sufficiently flat.
It seems naively that such tuning knows no bounds,
since the flatter the potential is, the longer the inflation lasts.
How flat is the inflaton potential?
Is it almost completely flat?

The minimal requirement of inflationary selection
\cite{Izaw}
is that the potential is so flat as
to induce sizable inflation. Observationally,
the potential of the primordial inflation
should be flat enough to realize several tens of $e$-folds.

As a possible further fine tuning, we here consider the parameter
tuning to realize eternal inflation.
If the fine-tuned parameter $\lambda$ is exceedingly small, the
slow-roll parameters
near $\varphi \simeq a/\lambda$ are extremely tiny. In such
a case, when $\varphi$ is very close to $a/\lambda$,
quantum effects dominate
field fluctuations and keep the system in a de Sitter
background effectively. This eternal inflation regime exists if quantum variance
$\Delta\varphi_{q} \simeq H/2\pi$ in the field $\varphi$ during the Hubble time $H^{-1}$ is larger than
the corresponding classical change $\Delta\varphi_{c}\simeq|-V'/3H^2|$,
where the Hubble scale $H\simeq \sqrt{V/3}$.

We now present the quantitative condition that a power-law potential
supports eternal phase
at $\v=a/\lambda$.
The quantum fluctuations are given by
\begin{eqnarray}
\Delta \varphi_q \simeq {1 \over 2\pi}\sqrt{V(a/\lambda) \over 3}
\simeq {1 \over 2\pi\sqrt{6n}}v^{2-n}a^n,
\end{eqnarray}
On the other hand, the slow-roll during the Hubble time
is given by
\begin{eqnarray}
\Delta \varphi_c \simeq {V'(a/\lambda) \over V(a/\lambda)}
\simeq {2n\lambda \over a}.
\end{eqnarray}
The slow-roll tends to be compensated by the
quantum fluctuations for
$\D \v_q \gsim \D \v_c$,
that is,
\beq
 {1 \over 2\pi\sqrt{6n}}v^{2-n}a^n \gsim {2n\lambda \over a}.
\eeq

The marginal inflation
is realized when this condition is
marginally satisfied for the $\phi$ upper bound $a$ of order one
to induce inflation.
Hence
the marginal or critical value of $\lambda$ to realize eternal inflation
in the present setup is obtained through
$\Delta \varphi_q \simeq \Delta \varphi_c$
as
\begin{eqnarray}
\lambda_{\rm marg} \simeq {1 \over 4\sqrt{6}\pi n^{3 \over 2}}v^{2-n}a^{n+1}.
\label{marg}
\end{eqnarray}

The scales of physical phenomena can be order of magnitude different
from one another such as QCD and electroweak scales. 
It is possible that the largest scale
in the effective field
theory, where the massive degrees of freedom as `heavy' as the
Planck scale have already been `integrated out',
is GUT scale.
That is, the highest scale (less than the Planck scale) to be dealt with might be GUT
scale of order $10^{-2}$.

Let us suppose that the scale in the $\phi$ potential term originates from this
largest physical scale.
Then the term $\phi^{4-2n}v^{2n}$
gives energy order of magnitude smaller than the Planck scale for
$|\phi| < a \lsim 1$, 
which is well-described in the framework of effective field theory.

To be specific, we here set the GUT-motivated `mass' scale
$v \simeq 1.5 \times 10^{-2}$ with $a \simeq 1$
as sample parameters.
Then, by means of Eq.(\ref{marg}),
$\lambda_{\rm marg} \simeq 1.7 \times 10^{-4}, 4.9 \times 10^{-4},
2.2 \times 10^{-3}$
for $n=0.5, 1, 1.5$, respectively.
These result in the amplitudes of the density fluctuations shown in Eq.(\ref{density})
as $\delta \rho/\rho \simeq (2.6-3.0) \times 10^{-5}, (1.8-2.2) \times
10^{-5}, (1.0-1.3) \times 10^{-4}$ for $n=0.5, 1, 1.5$, respectively,
under ${\cal N}=50-60$.
The obtained values happen to fall in the realistic regime (with slight
modification of the factor in the scale $v$, if necessary).

Namely, the considered tuning allows the primordial inflation
to be marginally eternal in the present setup.
The original GUT scale $v \sim 10^{-2}$ in the term $v^{4-2n}\phi^{2n}$
yields the physical scale in the rescaled
term $\lambda^{2n}v^{4-2n}\varphi^{2n}$ of intermediate scale
$\lambda^{2n/4-2n}v \sim 10^{-10}-10^{-3}$ for $\lambda \sim 10^{-3}$
and $n=0.5-1.5$,
which is appropriate for the primordial inflation.

Note that the fine tuning $\lambda \sim 10^{-3}$ is chosen
not by the GUT scale requirement $v \sim 10^{-2}$ under the observed
$\delta \rho/\rho$
but for marginal inflation.
The choice $\lambda \sim 10^{-1}-10^{-2}$ for non-eternal inflation
would be sufficient for primordial inflation, whereas
$\lambda < 10^{-4}$ would result in eternal inflation with over-tuning.

\section{A two-term case}

Let us make a remark on the case with two or more terms in the
potential.
We here adopt the simplest two-term example for presentation. 
The Lagrangian is given by
\begin{eqnarray}
{\cal L}={1 \over 2\lambda^2}\partial_\mu \phi \partial^\mu \phi
 - {1 \over 2}v^{2} \phi^{2} - {1 \over 4} \phi^{4}.
\end{eqnarray}
where the real scalar field $\phi$ takes values in the range $|\phi|<a$
as the effective theory description is applicable.
The constants $v$ and $a$ are again of order one or less
in the Planck unit
and the overall coupling $\lambda$ is the parameter to be tuned
for inflation within the regime $|\phi|<a$, as is the case for the
previous single term example.

We can define a rescaled field $\varphi=\phi / \lambda$ to make
the kinetic term for the field $\varphi$ canonical
and then its potential is given by
\begin{eqnarray}
V(\varphi)={1 \over 2}\lambda^{2} v^{2} \varphi^{2}
+{1 \over 4}\lambda^{4} \varphi^{4},
\end{eqnarray}
with $|\varphi|<a / \lambda$.

The critical coupling for marginal inflation is determined as
\begin{eqnarray}
\lambda_{\rm marg} \simeq {1 \over 16\sqrt{3}\pi}a^{3},
\end{eqnarray}
mainly due to the effect of the higher-order $\phi^4$ term.
If $a \simeq 0.5$, then $\lambda_{\rm marg} \simeq 10^{-3}$.

The observational value $\delta \rho / \rho \simeq 10^{-5}$
is dominantly realized for $v \simeq 10^{-2}$ by the $\varphi^2$ term.
More generally, lower-order terms tend to dominate observed density
fluctuations, while eternal inflation is mainly controlled by higher-order terms.

The $\lambda^2 v^2 \varphi^2/2$ and $\lambda^4 \varphi^4/4$ terms are comparable in size
around $\varphi \simeq 10\sqrt{2}$, which corresponds to
${\cal N} \simeq 50$ for $\varphi^2$ inflation. This implies that we have a possibility 
in realistic parameter choice for primordial inflation such that
the effects of the $\varphi^4$ correction to the $\varphi^2$ inflation
may be barely seen in the present cosmological observations.
For an analysis of inflationary dynamics under such circumstances, see polynomial
chaotic inflation investigated in
Ref.\cite{Nak}
albeit it deals with supergravity.
We conclude that 
such deviation from the single-term potential
may be detectable at the present epoch of our universe
if the primordial inflation is marginal.

\section{Conclusion}

We have investigated possible parameter choices
for primordial inflation with simple potentials
such as power-law and two-term potentials.
We examined the amount of parameter tuning
to make the slow-roll inflation eternal.
In particular, we adopted the critical coupling for marginally eternal inflation.

It turned out that potentials with mass scale of grand unification
provide, after tuning, consistent results to experimental data on
cosmological density fluctuations so that
the considered tuning is allowed observationally.
Moreover, the two-term case implies that the interplay
between effects of the two terms may be barely visible around the present
horizon scale of our universe.
In any case, our setup allows the primordial inflation to be marginally
eternal in contrast to the supersymmetric Higgs inflation
examined in a previous paper and thus provides a concrete example of
marginal primordial inflation. 

We note that the power-law and two-term potentials
provide just simplest examples which may describe
a portion of landscape of vacua.
Obviously, more involved setups such as plateau potentials
to realize small-field inflation should also be considered.
It is expected that future observations discriminate realistic
inflation among various possible ones and give us information on
parameter choice of nature in primordial inflation.
We hope that such knowledge eventually provide a clue
to reveal how the laws of nature, in particular, model parameters
in particle physics,
are chosen among possible various theories.

\section*{Acknowledgments}

We would like to acknowledge T.T.~Yanagida for valuable discussions and comments.
This work was supported by
World Premier International Research Center Initiative
(WPI Initiative), MEXT, Japan.

\end{document}